# SPONTANEOUS VIOLATION OF MIRROR SYMMETRY

*Igor T. Dyatlov[1]*
*Petersburg Institute of Nuclear Physics*
*Scientific Research Center "Kurchatov Institute", Gatchina, Russia*

A symmetry violation model is considered for a system that can spontaneously choose between identical states which differ from each other only in weak properties ($R \leftrightarrow L$). Such mirror symmetry allows reproduction of observed qualitative properties of quark and lepton mixing matrices. The lepton mixing matrix evidences in this case in favor of the inverse mass spectrum and the Dirac nature of SM neutrino. Notwithstanding the Dirac properties of neutrino, an exchange of lepton numbers such as $e^- + \mu^+ \rightarrow e^+ + \mu^-$ is possible but with only leptons participating in the process.

## 1. Introduction

The existence of mirror generations, which differ from Standard Model (SM) particles only in mass and weak properties, was first suggested by Lee and Yang [1] to explain unacceptable consequences of parity violation. We are talking about the possibility to distinguish physically the right-handed (R) and left-handed (L) coordinate systems.

Attempts to avoid the parity paradox produced a few hypotheses on mirror particles, the mirror world, which offered a multitude of interpretations (see review articles [2, 3] and a more recent article [4]). Previous papers [5, 6] by the author showed that mirror states of quarks and leptons participating in the formation of SM fermion mass matrices also allowed reproduction of the main qualitative properties of both weak mixing matrices (WMM) observed. The explanation why the two matrices were so different from each other was also naturally provided.

Then, during mass matrix production, SM particles should pass as though through the intermediate state of a heavy mirror analog. This is not just an additional condition imposed on the scheme. The very notion of "mirror symmetry" can be formulated in such a way that the transition through the intermediate state can become an important consequence of mirror symmetry violation in the mass fermion system.

In this case, the observed mass hierarchy of SM Dirac fermions, that is, quarks and charged leptons, and invariance with respect to the weak isospin *SU(2)* are essentially the only conditions necessary for the production of experimentally known structures and qualitative properties in both WMMs [7]. Both conditions are considered here as postulates of SM. Then, we will have reproduced the hierarchy of Cabibbo-Kobayashi-Maskawa (CKM) matrix elements, associated

Email: dyatlov@thd.pnpi.spb.ru

with the hierarchy of quark masses, and also obtain a completely different structure for the lepton WMM which will lack hierarchy and will be independent of mass spectrum.

It is noteworthy that, despite the participation of Majorana terms in the neutrino problem, the observed properties of lepton WMMs, when produced by the "mirror mechanism", can be achieved only if neutrinos are of the Dirac nature and have the so-called inverse mass spectrum [7, 8]. Moreover, the mirror scenario offers new argumentation, in addition to the see-saw mechanism [8, 9], that supports the exceptional smallness of neutrino masses.

Mirror generations are different from SM particles as a result of the violation of the original mirror symmetry. In [5, 6], this violation was introduced without determining its origin. In the present paper, this violation is assumed to have a spontaneous character and its possible mechanisms and implications are investigated. Concurrent with mass production, the weak isospin *SU(2)* group is violated. Therefore, a procedure similar to the SM mechanism can apparently be developed and use suitable Higgs scalars to achieve required vacuum averages.

Section 2 discusses the model of mirror symmetry and its violation as required in [5, 6]. This model differs from mirror-symmetrical schemes proposed by other authors and is believed to approach the scenario proposed by Lee and Yang most closely. Symmetry violation in this case results from the existence in the symmetrical system of alternative vacuum states that differ from each other only in the *R*- and *L*-properties.

Section 3 describes a specific model of mirror (and *SU(2)*) symmetry violation. It appears that the character of the chosen system (Section 2) defines the properties of the scalars participating in symmetry violation. Some implications of the mechanisms discussed are outlined in Conclusions (Section 4). The Dirac character of neutrinos, which precludes the double *β*-decay, does not prevent lepton number changes such as $e^- + \mu^+ \to \mu^- + e^+$ in our scenario but limit them to processes in which only leptons participate.

The use of scalars and their vacuum averages for symmetry violation (that is, mechanisms similar to SM) means inclusion in the system of strong Yukawa coupling. This problem has already been discussed earlier in connection with a very large mass of the *t*-quark [10]. The presence of even larger masses of mirror particles makes non-perturbative Yukawa coupling inevitable. This circumstance imposes significant limitations on any further discussion of mirror world effects on processes of accessible energies.

## 2. Mirror Symmetry

In [5, 6], the concept of mirror symmetry is based on the complete identity of the *R*- and *L*-systems prior to symmetry violation. Such understanding differs from other interpretations of mirror systems [2-4] whereby the *R*- and *L*-systems have different interactions, different representations of the various groups, different *R*, *L* vector bosons and even photons. In our opinion, identical systems correspond best to the original idea of Lee and Yang [1].

In the mirror-symmetrical system, the Lagrangian depends only on full Dirac operators, that is, doublets and singlets of the weak isospin $T_W$:

$$\Psi_{LR} = \psi_L + \Psi_R \ (T_W = 1/2), \quad \Psi_{RL} = \psi_R + \Psi_L \ (T_W = 0), \tag{1}$$

In (1) we omitted all other quantum numbers: $\bar{u}$ (up) and $\bar{d}$ (down) flavors, generation indices $n, n'$ = 1, 2, 3, etc. Eq. (1) shows massive Dirac fermions.

Mirror transformation is assumed to be an identical operation that is known a priori to preserve the Lagrangian invariant:

$$R \leftrightarrow L, \quad \psi \leftrightarrow \Psi. \tag{2}$$

The kinetic part and any gauge interaction in SM are written in terms of the operators (Eq.(1)) and are automatically separated into $\Psi$- and $\psi$-parts. Weak interaction is produced by the vector current of $\Psi_{LR}$ doublets. Mirror-symmetrical interactions with scalars, which define the properties of mass spectra, are discussed in Section 3, while the state masses (Eq.(1)) themselves are included directly in the mirror-symmetrical Lagrangian [5, 6]:

$$\mathcal{L}_0 = A\bar{\Psi}_{LR}\Psi_{LR} + B\bar{\Psi}_{RL}\Psi_{RL} + c.c., \tag{3}$$

where A and B are 3 x 3 matrices for generation indices. These terms are *SU(2)*-invariant, which means that:

$$A^{(\bar{u})} \equiv A^{(\bar{d})}, \quad B^{(\bar{u})} \neq B^{(\bar{d})}. \tag{4}$$

The mechanism of symmetry violation (2) is designed in much the same way as the scenario of SM.

The task of introducing $\psi$ and $\Psi$ asymmetry while maintaining the complete identity of the other properties can be achieved only if the system has two sets of states that differ from each other only in their *R*- and *L*-properties. There should be two possible ground states (vacuums) with different weak properties to satisfy this requirement. This means that, upon breaking of the $\psi \leftrightarrow \Psi$ symmetry, the weak current in one of the states, $|L\rangle$, will be left-handed for light $\psi$ particles and right-handed for heavy $\Psi$. The other state, $|R\rangle$, will show the opposite weak current characteristics.

The system being discussed can produce two types of vacuum averages:

$$\langle R|\varphi_1|R\rangle \equiv \langle L|\varphi_1|L\rangle \tag{5}$$

Equality results from the identity of other properties and

$$|\langle R|\varphi_2|L\rangle| = \left|\langle L|\varphi_2^+|R\rangle\right|. \tag{6}$$

Apparently, the field $\varphi_1$ is a scalar and $\varphi_2$ is a pseudoscalar. At least two types of "Higgs" scalars (pseudoscalars) and corresponding Yukawa couplings, which generate $\psi$- and $\Psi$-particle masses, are required. Differences in their masses should result from the existence of the different minima of the scalar field potential $V(\varphi_1, \varphi_2)$.

Before we proceed to the construction of a specific model, let us discuss some essential characteristics (positive and negative) of the proposed approach to spontaneous violation mechanism using scalar vacuum averages.

The problem of strong Yukawa coupling becomes crucial here: very heavy fermions are an integral part of the system in general. Of course, we cannot solve this problem. Moreover, in the Lagrangian itself (see (10)), only heavy mirror $\Psi$ particles interact with the observed scalar *H*. We will, however, show that the mechanism of mass production (Section 3) will conserve in this case the usual perturbative interaction of SM's "light" Dirac fermions (at the existing masses and vacuum average $\eta \approx 246$ GeV) with the Higgs particle *H* (the coupling constant $h \sim m/\eta$). This is a direct consequence of system invariance with respect to the weak isospin *SU(2)* and does not require any further assumptions.

It appears that nature, having created SM, has probably separated the perturbative part out of the large, common system.[2] All coupling constants remain in the perturbative region. For this purpose, SM particles would have to be broken from mirror states. No strong interconnections should exist between the two parts. Two important characteristics observed in SM make this possible:

1. Although the complete system—that is, SM particles + mirror generations (*L, R*-symmetrical system) – does not have chiral anomalies [11], its low-energy part (SM) is known to be devoid of anomalies in itself and is thus renormalizable and may be perturbative. Complex cancellation of anomalous quark and lepton contributions takes place. If low-energy contributions did not cancel out, this would indicate strong coupling between all tiers of the system which would preclude the required breaking of low states from high ones.

---

[2] Strong quark interactions do not play any role in the formation of fundamental masses. At high energies, they are practically absent.

2. The small mass of the Higgs boson ($m_H \approx 126$ GeV, [12]) not only maintains the perturbative unitarity in processes involving longitudinal vector bosons ($m_H < 1$ TeV) but also provides perturbative character of the self-action $\lambda$ of scalars ($V = \lambda \phi^4$, $m_H < 0.5$ TeV). Large masses of the boson $H$ would also lead to strong coupling with high-energy states.

Concluding this section, let us note that the system of mirror symmetry adopted by us provides an answer to the question that other scenarios [3] fail to answer—that is, are all mirror particles heavier than SM particles? The mechanism being discussed in this paper reproduces the WMM only if very heavy mirror fermions with masses much larger than the masses of SM particles participate in the process of SM mass matrix production.

Yet, the origin of the Dirac fermion mass hierarchy in SM or the actual causes of this phenomenon continue to remain uncertain (see [13]). Here we only mechanically transfer the problem to the unknown origin of the mirror particle hierarchy (inverse to SM).

## 3. Model of Mirror Symmetry Spontaneous Violation

In accordance with the conclusions in Section 2, let us consider the Yukawa couplings of mirror-symmetrical operators (1) with complex $\varphi_1$ (scalar) and $\varphi_2$ (pseudoscalar) isodoublets. In complete compliance with SM, we have:

$$\begin{aligned}\mathcal{L}' &= (h_1^{(\bar{u})})_n^{n'} \left( \bar{\Psi}_{LR}^n, \varphi_1^c \right) \Psi_{RL\,n'}^{(\bar{u})} + (h_2^{(\bar{u})})_n^{n'} \left( \bar{\Psi}_{LR}^n \gamma_5, \varphi_2^c \right) \Psi_{RL\,n'}^{(\bar{u})} + c.c. \;+ \\ &+ \left( h_1^{(\bar{d})} \right)_n^{n'} \left( \bar{\Psi}_{LR}^n, \varphi_1 \right) \Psi_{RL\,n'}^{(\bar{d})} + \left( h_2^{(\bar{d})} \right)_n^{n'} \left( \bar{\Psi}_{LR}^n \gamma_5, \varphi_2 \right) \Psi_{RL\,n'}^{(\bar{d})} + c.c.\end{aligned} \quad (7)$$

In Eq. (7) all state indices characterizing the system being used—that is, flavor $f = \overline{u, d}$, generation indices $n, n' = 1, 2, 3$—are written out. Round parentheses $(\bar{\Psi}, \varphi)$ denote isodoublet scalar products, $\varphi^c = i\sigma_y \varphi^+$. Further on, to avoid cumbersome formulae, we will omit most of the symbols.[3]

Let us substitute operators (1) in Eq. (7) to obtain:

$$\begin{aligned}\mathcal{L}' &= h_1 \Big[ (\bar{\Psi}_R, \varphi_1) \Psi_L + (\bar{\psi}_L \varphi_1) \psi_R \Big] + h_2 \Big[ -(\bar{\Psi}_R, \varphi_2) \Psi_L + (\bar{\psi}_L, \varphi_2) \psi_R \Big] + \cdots = \\ &= \left( \bar{\Psi}_R, [h_1 \varphi_1 - h_2 \varphi_2] \right) \Psi_L + \left( \bar{\psi}_L, [h_1 \varphi_1 + h_2 \varphi_2] \right) \psi_R + \cdots\end{aligned} \quad (8)$$

Upon mirror symmetry violation, the identity of the $\Psi$ and $\psi$ systems is only achievable if there is direct relation between the matrices $h_1$ and $h_2$.

---

[3] Eq. (7) can be written in a diagonal form in terms of generation indices; at that, Eq. (18) must always be non-diagonal.

Let us consider the simplest case of such a connection:

$$h_1 \equiv h_2 = h, \tag{9}$$

which demonstrates more readily the qualitative essence of the mechanism (we can assume the matrix $h$ to be diagonal, without losing generality). Indeed (see Eq. (12)), in this case, "the other world" is completely similar (apart from the weak properties: $R \leftrightarrow L$) to SM.

Then the Langrangian (8) will have the form:

$$\begin{aligned} \mathcal{L}' &= h(\bar{\Psi}_R, \Phi_1)\Psi_L + h(\bar{\psi}_L, \Phi_2)\psi_R + \cdots \\ \Phi_1 &= \varphi_1 - \varphi_2, \quad \Phi_2 = \varphi_1 + \varphi_2. \end{aligned} \tag{10}$$

In the mirror-symmetrical world, the operators $\Phi_1$ and $\Phi_2$ do not have well-defined parity. Under broken mirror symmetry, however, this does not add any new features to the system: the same situation occurs in SM for the ordinary Higgs scalar [14]. Similar to SM, the same bosons $\Phi_1$ and $\Phi_2$ produce a Yukawa coupling for both quarks and leptons. This is necessary, to avoid an increased number of Goldstone phases of boson $SU(2)$ doublets. Three phases of one of the operators $\Phi_1$, $\Phi_2$ achieving a vacuum average are sufficient to increase the masses of all three weak vector bosons $W_\mu$.

Let us take $\Phi_1, \Phi_2$—a symmetrical expression constructed entirely analogous to SM—as a potential for $V(\Phi_1,\Phi_2)$ scalars:

$$V(\Phi_1, \Phi_2) = \kappa |\Phi_1|^2 |\Phi_2|^2 - \frac{\mu^2}{2}\left(|\Phi_1|^2 + |\Phi_2|^2\right) + \frac{\lambda}{4}\left(|\Phi_1|^4 + |\Phi_2|^4\right). \tag{11}$$

At large $\kappa$, the deepest minima $V$ divide the system both by $\Phi_1$ and $\Phi_2$ and by $\Psi, \psi$. For vacuum averages, we obtain from Eq. (11):

$$\langle \Phi_2 \rangle = 0, \quad \langle \Phi_1 \rangle^2 = \frac{\mu^2}{\lambda} = \eta^2, \tag{12a}$$

$$\langle \Phi_1 \rangle = 0, \quad \langle \Phi_2 \rangle^2 = \frac{\mu^2}{\lambda}. \tag{12b}$$

In an interesting scenario for SM (Eq. (12a)), the operator $\Phi_1$ is a system consisting of one neutral scalar $H$ (the Higgs boson of SM) and Goldstone phases which produce masses of $W$-bosons. It is noteworthy that the tree-level mass of the other boson, $\Phi_2$, equal to

$$M_{\Phi_2}^2 = \kappa \frac{\mu^2}{\lambda} - \frac{\mu^2}{2}, \tag{13}$$

can be made, at large $\kappa$, as large as desired:

$$\kappa\eta^2 \gg \frac{\mu^2}{2}. \tag{14}$$

The operator $\Phi_2$ produces four scalar states: two charged (±) and two neutral, as it occurs in the case of $K\overline{K}$ mesons.

Let us now discuss an unfortunate circumstance associated with the inevitable presence, in our scenario, of a nonperturbative Yukawa interaction. If $\Phi_1$ is identified with the Higgs scalar *H*, then the quantity $\langle\Phi_1\rangle = \eta$ is defined by the mass of the *W*-boson ($M_W = g_W\eta/2$, $\eta \simeq 246$ GeV). The large mass of $\Psi$ particles means therefore larger values of $h \simeq M_\Psi/\eta \gg 1$. This in fact halts further quantitative use of the proposed scheme. In particular, the influence of mirror quarks on Higgs boson production, which is known to involve heavy-quark triangle-diagram contributions [14] (see also Section 4), needs yet to be explained in quantitative terms.

At the same time, the interaction of the standard Higgs boson *H* with SM's light fermions, $\psi$, is consistent with what we have in SM. In fact, although the Higgs scalar $\Phi_1$ in Eq. (10) interacts directly only with mirror states $\Psi$, the diagonalization of mass matrices of the full system $\psi, \Psi$ results in eigenfunctions of massive states of the $\Psi_M$ and $\psi_\mu$ type (see Section 7 in [5]):

$$\begin{aligned}\Psi_M &= \Psi + O\left(\left(\frac{\mu_\psi}{M_\Psi}\right)^{1/2}\right)\psi, \\ \psi_\mu &= \psi + O\left(\left(\frac{\mu_\psi}{M_\Psi}\right)^{1/2}\right)\Psi.\end{aligned} \tag{15}$$

The right-hand side of Eq.(15) shows sums over generation indices. Then the interaction $h_\psi \overline{\psi}_\mu \psi_\mu \Phi_1$ will have a coupling constant typical of SM. From Eq.(10) and Eq.(15), we have:

$$h_\psi = h\left(\frac{\mu_\psi}{M_\Psi}\right)^{1/2}_1 \left(\frac{\mu_\psi}{M_\Psi}\right)^{1/2}_2 \sim \frac{M_\Psi}{\eta} \cdot \frac{\mu_\psi}{M_\Psi} \sim \frac{\mu_\psi}{\eta}. \tag{16}$$

Let us prove that this result is a direct consequence of weak *SU*(2) symmetry.

In fact, symmetry means that in the transverse part of the vector boson propagator:

$$\Delta_{\mu\nu} = \frac{g_{\mu\nu} - (q_\mu q_\nu)/q^2}{M_W^2 - q^2}$$

at $q^2 = 0$, the pole should be cancelled in the invariant gauge together with Goldstone contributions produced by the phases of the scalar $\Phi_1$. Cancellation should take place in contributions of any diagrams of the interaction of $\psi$ fermions with $W$ (Fig. 1), which is precisely the case owing to Eq.(15) and Eq.(16).

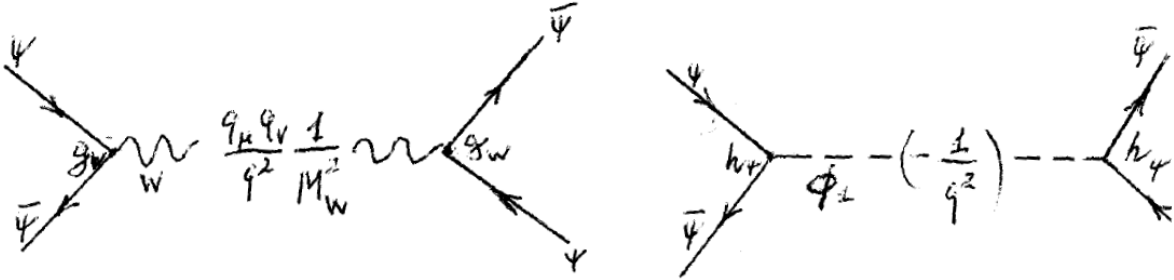

Fig. 1: Cancellation of the W-propagator poles and Goldstone boson in invariant gauge of the gauge theory, $M_W = g_W \eta/2$.

The Lagrangian (10) produces Dirac mass terms of mirror particles:

$$\mathcal{L}' = M_\Psi \bar{\Psi}_R \Psi_L + \cdots . \tag{17}$$

Along with the mass matrices of the mirror-symmetrical states $\Psi_{LR}$ and $\Psi_{RL}$ from the original Lagrangian (3), which, upon mirror symmetry breaking, make transitions $\Psi \leftrightarrow \psi$ [5, 6],

$$\mathcal{L}_0 = A\bar{\Psi}_{LR}\Psi_{LR} + B\bar{\Psi}_{RL}\Psi_{RL} + c.c. = A\bar{\psi}_L \Psi_R + B\bar{\psi}_R \Psi_L + c.c. \cdots , \tag{18}$$

we obtain a system of equations for mass matrices of $\psi$ particles. Under conditions specified in [5], this system reproduces the qualitative properties of the CKM matrix. These conditions are the observed hierarchy of quark masses and the consequences of *SU*(2) symmetry (4).

In describing the mechanism of neutrino mass production, one should suggest that, according to [6], Majorana mass terms will also be produced. This will, of course, require "purely leptonic" isoscalars. These do not have any influence on the mass of *W*.

For singlet charge-symmetrical states $\Psi_{RL}$ in the case of neutrino, Majorana terms may be produced similar to the procedure in Eq. (7)-(10) and Eq. (11)-(13). This procedure now requires use of two isoscalars, $\varphi'$ and $\varphi''$ (pseudoscalar). Identical operations result in the following expression (*C* is a charge conjugation matrix):

$$\begin{aligned}\mathcal{L}_\nu &= h_\mathcal{M} \Psi_{RL}^T C \Psi_{RL} \varphi' + h_\mathcal{M} \Psi_{RL}^T C \gamma_5 \Psi_{RL} \varphi'' + c.c. = \\ &= h_\mathcal{M} \Psi_L^T C \Psi_L \Phi' + h_\mathcal{M} \psi_R^T C \psi_R \Phi'' + c.c. \\ \Phi' &= \varphi' - \varphi'', \quad \Phi'' = \varphi' + \varphi''.\end{aligned} \tag{19}$$

Operations with isoscalars $\Phi'$ and $\Phi''$ are easy and obvious; their masses can be taken as large as desired ($\mathcal{M}'_L = h_\mathcal{M} \langle \Phi' \rangle$) while complex phases are not important and can be transposed to fermion operators. Eq.(19) is written out certainly only for the neutrino flavor. Similar to (11), two symmetrical vacuums can be constructed; the vacuum that interests us is $\langle\Phi'\rangle \neq 0, \langle\Phi''\rangle = 0$. The procedure in [6], however, requires that, for generation of the experimentally acceptable lepton WMM, the doublet part $\Psi_R$ of mirror neutrinos have a similar Majorana mass $|\mathcal{M}'_L|$. Moreover, $\mathcal{M}_R$ must have the opposite sign:

$$\mathcal{L}_\nu = \mathcal{M}_R \Psi_R^T C \Psi_R + c.c., \quad \mathcal{M}_R = -\mathcal{M}'_L. \tag{20}$$

The origin of the mechanism (20) contains a certain complication that has been mentioned numerous times in papers on Majorana masses and the see-saw mechanism [8, 9]. To preserve *SU*(2) invariance in the analog of Eq.(19) for isodoublets $\Psi_{LR}$ requires introduction of either isotriplet scalars $\Phi_{T=1}$ or non-renormalizable terms with the square of doublets $\Phi_1, \Phi_2$. Both options are by no means considered to be perfect.

We propose a different hypothesis and different argumentation for Eq. (20). Both continue to be only verbal, again due to the presence of the strong Yukawa coupling. In fact, Eq. (20) must appear in our scenario even in the absence of such terms in the Lagrangian itself. Solving the mass problem for higher order interactions will lead to Eq. (20). Fig.(2a) shows the simplest diagram reproducing $\mathcal{M}_R$. The line $\Phi_1$ in this diagram denotes the scalar Higgs particle *H*. Only such a neutral (real-valued) state can result in the array shown in Fig.(2), because $\langle\Phi'\rangle \neq 0$ appears only in the case of neutrino components $\Psi_L$ in Eq.(19), and only the neutral component of the lepton doublet $\Psi_R$ can be used in Fig.(2a). It is not possible to have a similar process involving $\Phi_2$ for $\psi_L$, because particles in $\Phi_2$, even in its neutral part, differ from antiparticles. Only radiation processes of the type shown in Fig.(2b), which violate lepton numbers, are possible.

Triangle diagrams in Fig.(2a), no matter how complex they are, lead to a logarithmic divergence whose contribution is equivalent to $\mathcal{M}_R$. This contribution needs to be somehow fixed. All other divergences that are present in the model occur naturally in the renormalizable theory and can be removed by means of redefinition of constants and introduction of counter-terms. The divergence in Fig.(2a) is the only divergence which cannot be removed due to the absence of adequate terms in the Lagrangian.

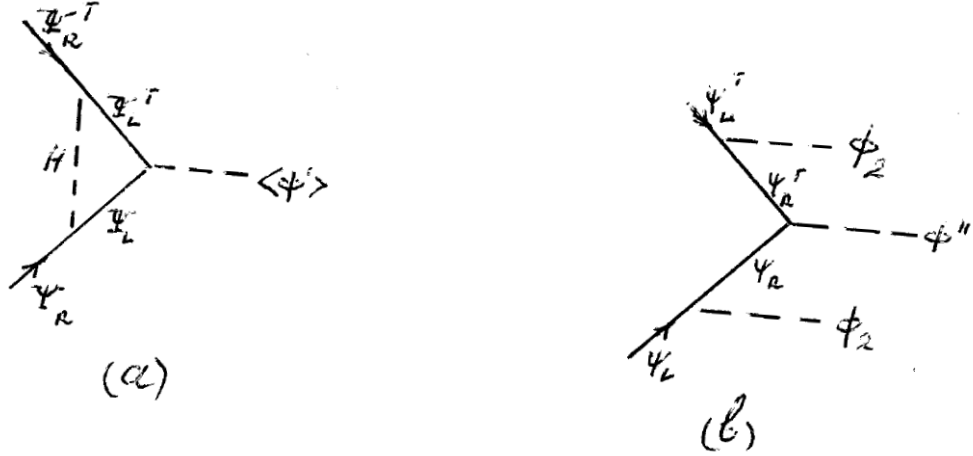

Fig. 2: Production of the mirror Majorana mass $M_R$ (Fig. 2a) and non-conservation of the lepton number in SM particles (Fig. 2b).

Energy considerations may influence the choice of the divergence value. If Majorana masses are very heavy—which is necessary in order to obtain very small $m_\nu$ [6]—then their contribution to energy (energy is independent of the fermion mass sign) will be very large:

$$\left| \mathcal{M}_R \Psi_R^T C \Psi_R + \mathcal{M}_L \Psi_L^T C \Psi_L \right| \tag{21}$$

This expression is minimal if the system is of the sort that $\Psi_R$ and $\Psi_L$ are only in the *L,R* symmetrical states and if $\mathcal{M}_R = -\mathcal{M}_L$. This occurs when neutrino is Dirac in nature, which permits to reproduce, as is shown in [6], the observed form of the neutrino WMM.

Based on the above, we believe that $\mathcal{M}_R$ may appear not spontaneously, through vacuum averages, but rather as a result of model dynamics, and the equality $\mathcal{M}_R = -\mathcal{M}_L$ depends on the energy minimum chosen.

## 4. Conclusion. Influence of Mirror World

Mirror quark and lepton generations may be widely spaced from accessible energies. For quark masses, one can only make an assessment based on the necessary condition imposed on the mechanism described in [5, 6]: mirror particles must be much heavier than SM particles and their mass hierarchies must be inverse to hierarchies of ordinary fermions. The mirror analog of the *t*-quark must have the smallest mass (see [5]).

A unit character of mirror particle production is possible, however, it is weakened by small mixings (15). The link with mirror physics through the Higgs scalar *H* brings under consideration the non-perturbative interaction whose role is currently unknown.

First of all this concerns the very process of the Higgs boson *H* production. Two gluon jets produce H: $gg \to H$, the process that played a major role in the discovery of the boson at LHC [12]. The triangle of heavy quarks that creates here the source of bosons does not depend, as is known, on the mass of the quarks that form it [14]. Mirror particles with large masses could provide an enormous contribution, approximately $16 = 4^2$ times larger than that provided by a single $t$-quark of SM.

For realistic evaluation, however, one has to calculate the sum of all triangle diagrams for the non-perturbative case of the $\bar{\Psi}\Psi H$ coupling, which has not been done.

The possibility of estimating mass values of unknown scalars $\Phi_2, \Phi$ and $\Phi'$ is even lower.

Heavy leptons and bosons may lead to additional contributions in processes of the $\mu \to e\gamma$ type. Mirror lepton corrections in the phenomena involving SM particles are weakened significantly because, as in the case of quarks, they undergo the earlier mentioned corrections in Eq.(15) for the coupling of mirror states with SM states. Boson contributions, again, require non-perturbative calculations.

The most interesting processes, in terms of potential observation, are those involving "charge exchange" of lepton numbers of different generations of the following type:

$$e^- + \mu^+ \longrightarrow e^+ + \mu^-. \tag{22}$$

In our model, these processes occur as a result of the Majorana interaction of leptons (19) and the production of heavy bosons, as shown in Fig.(2b). The schematic presentation of the lepton charge exchange mechanism is shown in Fig.(3). One should remember that the Yukawa couplings of the boson $\Phi_2$ (Eq.(10)) in our model are equal to the constants of coupling of the Higgs bosons $\Phi_1$ with mirror leptons—that is, they can be non-perturbative. In our instance, this type of phenomena are possible exclusively in lepton processes, because all neutrino are Dirac in nature.

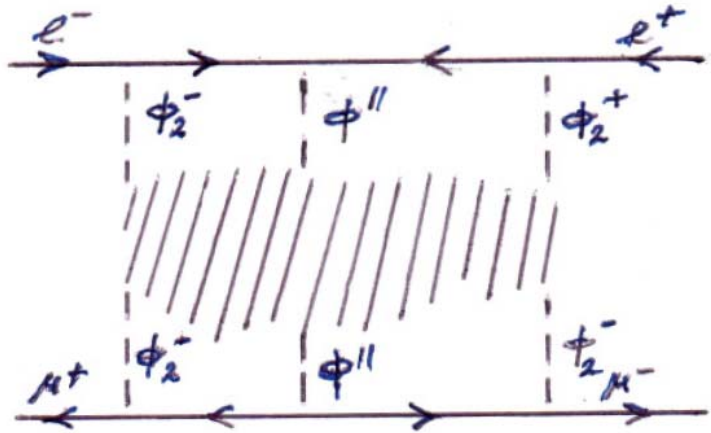

Fig. 3: Lepton number exchange mechanism. Hatched area shows the non-perturbative coupling of the boson $\Phi_2$.

The process (22) is also possible when SM neutrino is Majorana in nature, independently of our scenario involving mirror particles. In this case, the cross-section of neutrino is tiny for observed neutrino masses and even for $m_v \approx M_Z/2$ it is $\lesssim 10^{-7}$ times the weak process cross-section $d\sigma_W$; for example, $v + e \to v + e$. A rough estimate of the contribution in Fig.(3) based on geometry (all that falls within the domain of mirror masses entirely strongly interacts) and the maximum possible size of the domain ($E_{min} \sim$ several TeV) gives the value $d\sigma \lesssim 10^{-4\text{-}5} d\sigma_W$; however, at $E > M_W$ the contribution may exceed this estimate significantly. Of course, this estimate needs a more profound consideration. It should also be mentioned that $E_{min}$ is defined here not by the masses of mirror leptons but by unknown bosons, which may lead to other (even smaller) energy scales.

The process (22) appears to be convenient for investigation. It does not have a threshold, its cross-section increases over a long time (up to $E_{min}$) with energy (much longer than low energy weak processes), and $\mu$-meson beams are quite accessible.

The author is grateful to Ya. I. Azimov and M.G. Ryskin for their interest in this work and for useful discussions. This work was funded by grant RSF No. 14-92-0028.